\documentclass[amsmath,amssymb,aps,prd,showpacs,nofootinbib]{revtex4}

\usepackage{graphicx}
\begin{document}

\title{Strong decays with the boost-corrected wave functions}


 \author{\firstname{Yu.A.}~\surname{Simonov}}
\email{simonov@itep.ru} \affiliation{NRC ``Kurchatov Institute'' -- ITEP, B. Cheremushkinskaya 25, Moscow, 117259, Russia}


\newcommand{\beq}{\begin{eqnarray}}
 \newcommand{\eeq}{\end{eqnarray}}
\newcommand{\be}{\begin{equation}}
 \newcommand{\ee}{\end{equation}}

 \def\la{\mathrel{\mathpalette\fun <}}
\def\ga{\mathrel{\mathpalette\fun >}}
\def\fun#1#2{\lower3.6pt\vbox{\baselineskip0pt\lineskip.9pt
\ialign{$\mathsurround=0pt#1\hfil ##\hfil$\crcr#2\crcr\sim\crcr}}}
\newcommand{\veX}{\mbox{\boldmath${\rm X}$}}
\newcommand{{\SD}}{\rm SD}
\newcommand{\pp}{\prime\prime}
\newcommand{{\Mc}}{\mathcal{M}}
\newcommand{\veY}{\mbox{\boldmath${\rm Y}$}}
\newcommand{\vex}{\mbox{\boldmath${\rm x}$}}
\newcommand{\vey}{\mbox{\boldmath${\rm y}$}}
\newcommand{\ver}{\mbox{\boldmath${\rm r}$}}
\newcommand{\vesig}{\mbox{\boldmath${\rm \sigma}$}}
\newcommand{\vedelta}{\mbox{\boldmath${\rm \delta}$}}
\newcommand{\veP}{\mbox{\boldmath${\rm P}$}}
\newcommand{\veA}{\mbox{\boldmath${\rm A}$}}
\newcommand{\vep}{\mbox{\boldmath${\rm p}$}}
\newcommand{\veq}{\mbox{\boldmath${\rm q}$}}
\newcommand{\veQ}{\mbox{\boldmath${\rm Q}$}}
\newcommand{\vez}{\mbox{\boldmath${\rm z}$}}
\newcommand{\veS}{\mbox{\boldmath${\rm S}$}}
\newcommand{\veL}{\mbox{\boldmath${\rm L}$}}
\newcommand{\veR}{\mbox{\boldmath${\rm R}$}}
\newcommand{\ves}{\mbox{\boldmath${\rm s}$}}
\newcommand{\vek}{\mbox{\boldmath${\rm k}$}}
\newcommand{\ven}{\mbox{\boldmath${\rm n}$}}
\newcommand{\veu}{\mbox{\boldmath${\rm u}$}}
\newcommand{\vev}{\mbox{\boldmath${\rm v}$}}
\newcommand{\veh}{\mbox{\boldmath${\rm h}$}}
\newcommand{\vew}{\mbox{\boldmath${\rm w}$}}
\newcommand{\verho}{\mbox{\boldmath${\rm \rho}$}}
\newcommand{\vexi}{\mbox{\boldmath${\rm \xi}$}}
\newcommand{\veta}{\mbox{\boldmath${\rm \eta}$}}
\newcommand{\veB}{\mbox{\boldmath${\rm B}$}}
\newcommand{\veH}{\mbox{\boldmath${\rm H}$}}
\newcommand{\veE}{\mbox{\boldmath${\rm E}$}}
\newcommand{\veJ}{\mbox{\boldmath${\rm J}$}}
\newcommand{\veal}{\mbox{\boldmath${\rm \alpha}$}}
\newcommand{\vepi}{\mbox{\boldmath${\rm \pi}$}}
\newcommand{\vegam}{\mbox{\boldmath${\rm \gamma}$}}
\newcommand{\vepar}{\mbox{\boldmath${\rm \partial}$}}
\newcommand{\llan}{\langle\langle}
\newcommand{\rran}{\rangle\rangle}
\newcommand{\lan}{\langle}
\newcommand{\ran}{\rangle}

\begin{abstract}
 Strong decay probabilities are calculated using the Lorentz contracted wave functions of decay products, determined in the arbitrary dynamical scheme with the instantaneous interaction. It is shown that the decay width acquires an additional factor, defined by the contraction coefficient $C_m(s)$, which for the two-body equal mass decays is  $C^2_m(s)= 4m^2/s$ , $s= E^2$. The resulting decay widths are compared to  experimental data, where, in particular the $\rho(770),\rho(1450) $ decay data, require an additional $1/s$ dependence of the width to fit the data. Important consequences for the dynamics of hadron decays and scattering are shortly discussed.
\end{abstract}
\maketitle
 \section{Introduction}

In the theory of hadron transformations, e.g. the hadron decays into other hadrons, or the hadron exchanges between hadrons, or the hadron form factors,  one is met
with the matrix elements, which include overlap integrals of the hadron wave functions. Typically these wave functions correspond to hadrons with different
momenta. For example, the theory of a hadron decay into other hadrons or hadrons plus the elementary objects include the overlap integrals of the hadron
wave functions, which correspond to the instantaneous images of these hadrons, and the question arises whether the Lorentz contraction of moving hadrons is
taken into account or not. For this purpose one needs to describe the motion and interaction of extended objects and for that one has to know behavior of the Green's
functions and the wave functions of extended objects under the applied boost, e.g. to know how the velocity $\vev$  of the system affects the hadron wave function.

As an example one can consider  the hadron decay matrix element of the process $h\rightarrow h_1 + h_2$, e.g. $\rho\rightarrow \pi+\pi$, where the pions move with
high velocity and therefore their wave functions enter in the strong decay matrix element in the Lorentz transformed way.

It is a purpose of present paper to derive the behavior of the hadron wave functions in the moving system and calculate the
resulting behavior of the hadron decay matrix element. As it is known \cite{1}, in the  relativistic  field theory the general formalism can be constructed in three different ways:
\begin{description}
\item{1)} the instant form,
\item{2)} the point form and
\item{3)} the light front form.
\end{description}

In the instant form (IF) the wave function of any nonlocal object, consisting of several elements, can be defined at one moment of time and the frame (boost) dependence
is dynamically generated in connection with Hamiltonian. In the literature different approaches have been developed for the practical realization of this problem,
e.g. the canonical formalism in \cite{2,2*}, the analysis of the operator matrix elements between wave functions and form factors \cite{3}.
The analysis of different forms of interaction in the IF was done in \cite{3*,4*}, where it was shown that the
relativistic conditions can be satisfied with a large class of interactions, including the standard nonrelativistic potentials. As will be seen below it is in the IF formalism one can introduce and study the reaction amplitudes, like hadron decays, which contain the wave functions of participating objects.
In what follows we shall be interested first of all in the strong decay processes which contain explicitly the integrals of the wave functions
of hadrons and we shall study how those are deformed due the boost acquired in the decay process.
As it is clear, this is possible only in the IF of relativistic wave functions and we shall always consider this form.

 As it is,
the theory of the frame dependence of the Green's functions of any nonlocal objects  is closely related to the properties of the interaction terms in the Lagrangian,
and one must envisage the instantaneous interaction for the process, in particular, confinement for the strong interaction and the Coulomb force in QED. The
dynamical studies in this direction have been done  recently, in Refs.~\cite{4,5,6} in several examples of systems. Later on, in \cite{7} the properties of the spectrum
and the wave functions in the moving system were studied in the framework of  the relativistic path integral formalism \cite{8,9,10,11}.  This method essentially
exploits the universality and the Lorentz invariance of the Wilson-loop form of the interaction, which produces both confinement and the gluon-exchange interaction in QCD.
Moreover, in this formalism the Hamiltonian $H$ with the instantaneous interaction between the quarks in QCD (called the relativistic string Hamiltonian (RSH)) and charged particles
in QED was derived, where the known defects of the Bethe--Salpeter approach are missing. In \cite{7} it was shown that the eigenvalues and the wave functions, defined by the RSH,
transform in the moving system in accordance with the Lorentz rules. Indeed, using the invariance law under the Lorenz transformations \cite{12,13},
\be
\rho(\vex,t)dV = {\rm invariant},~~\label{eq.01}
\ee
where $\rho(\vex,t)$ is the density, associated with the wave function $\psi_n(\vex,t)$,
\be
\rho_n(\vex,t) = \frac{1}{2i} \left(\psi_n \frac{\partial \psi_n^+}{\partial t}  - \psi_n^+ \frac{\partial \psi_n}{\partial t}\right)
= E_n |\psi_n(\vex,t)|^2, ~~\label{eq.02}
\ee
and $dV=d\vex_{\bot} dx_{\|}$. One can use the standard transformations,
\be
L_{\rm P}dx_{\|} \rightarrow dx_{\|} \sqrt{1 - \vev^2}, ~~ L_{\rm P} E_n \rightarrow \frac{E_n}{\sqrt{1-\vev^2}},
\label{eq.03}
\ee
to ensure the invariance of  (\ref{eq.01}). In its turn the invariance law implies that in the wave function $\psi(\vex,t)=\exp(-iE_nt)\varphi_n(\vex)$ the function $\varphi_n(\vex)$ is deformed in the moving system,
\be
L_{\rm P}\varphi_n(\vex_\bot, x_{\|}) = \varphi_n\left(\vex_\bot, \frac{x_{\|}}{\sqrt{1-\vev^2}}\right),
\label{eq.04}
\ee
and can be normalized as
\be
\int E_n |\varphi_n^{(v)}(\vex)|^2 dV_v = 1 = \int M_0^{(0)} |\varphi_n^{(0)}(\vex)|^2 dV_0,
\label{eq.05}
\ee
where the subscripts $(v)$ and $(0)$ refer to the moving and the rest frames. One of the immediate consequences from the
Eqs.~(\ref{eq.03}) and (\ref{eq.04}) is the property of the boosted Fourier component of the wave function:
\be
\varphi_n^{(v)}(\veq) = \int \varphi_n^{(v)}(\ver) \exp(i\veq\ver) d\ver = C_0 \varphi_n^{(0)}(\veq_\bot, q_{\|}\sqrt{1-v^2}),
\label{eq.06}
\ee
where $C_0 = \sqrt{1-v^2} = \frac{M_0}{\sqrt{M_0^2 + \veP^2}}$.

The equations (\ref{eq.01}) -- (\ref{eq.06}) and in particular (\ref{eq.06}), formulated in Ref.~\cite{7}, have been  the basic elements of the analysis of the meson form factors
in Ref.~\cite{14}, where it was shown that the Lorentz contraction
of the hadron wave functions creates a basically different behavior of the form factors as the functions of $Q^2$, such that arguments of the wave functions are never occur
in the asymptotically large momenta region. In the concrete examples of the
pion and kaon form factors the agreement with data was obtained with simple Gaussian wave functions in the whole region of $Q^2$ \cite{14}. A similar situation holds for the proton and neutron form factors \cite{15}.

It is the purpose of the present paper to study the behavior of the hadron strong decay matrix elements using the
Lorentz contracted wave functions of decay products. To this end we need explicit expressions of the decay matrix elements in terms of these wave functions.

In section 2 we shall write the expressions for the meson decay matrix elements in the rest frame of the decaying  meson. For this purpose we use the relativistic theory of string breaking
\cite{16}, \cite{17}, which is an extension and the relativistic version of the well-elaborated strong decay formalism, based
on the original $^3P_0$ model \cite{18} and its flux-tube modification \cite{19}(for the analysis of the model see \cite{20} and the reviews in \cite{21}). In what
follows we shall need also the modified forms of the string breaking matrix elements, see e.g. \cite{17}. In the case of the chiral mesons as the decay products
we shall be using the formalism, called the Chiral Decay Mechanism (CDM), described in \cite{22,23,24}. The technique of the Fock--Feynman--Schwinger representation (FFSR)
\cite{9,10,11} allows to represent the results in a simple form, which can be compared to experimental and lattice data in section 3. In section 4 we discuss the consequences
and extrapolations of our results, as well as possible implications of the Lorentz contraction for other hadron decays. The concluding section contains a summary of results and discussion.

\section{Definition of the decay matrix element through the hadron wave functions }

The standard local form of the decay Lagrangians e.g. $\rho $ decay is $ L_\rho = g e_{ijk} \rho^i_\mu \pi^j
\partial_\mu \pi_k$ with the resulting width $\Gamma_\rho(E)= \frac{g^2 p^3(E)}{6\pi m_\rho^2}$, where $p(E)$ with
$E$ around the $\rho$ mass is $p(E)= \sqrt{\frac{E^2}{4} - m_\pi^2}$. As will be discussed
in the next sections the experimental data display a different behavior at large $E$ and it will be our aim
to explain this difference. To this end one must transform from the effective local form above to the detailed
form containing the hadron wave functions.

We start with the simplest form of the $^3P_0$ model as a interaction Hamiltonian
 \be H_I = g \int {d^3x \bar\psi \psi}, \label{0.7} \ee
 where $g = 2 m_q \gamma$, and $\gamma$ is a phenomenological parameter. The relativistic form, obtained in \cite{17},  can be written as
 \be S_{eff} = \int { d^4 x \bar\psi(x) M(x) \psi(x)}, \label{0.8} \ee
 where $M(x) = \sigma (|\vex -\vex_Q| + |\vex - \vex_{\bar Q}|)$. Here $\vex$ is the string breaking point between the quarks $Q$ and $\bar Q$.
In the momentum space one obtains, as in \cite{17,18} for the decay of the hadron $1$ into hadrons $2,3$
\be J_{123}(\vep) = y_{123} \int{\frac{d^3q}{2\pi^3} \Psi_1(\vep , \veq) M(q) \psi_2(\veq) \psi_3(\veq)}.
\label{0.8a} \ee

Here $\vep,-\vep$ are the momenta of the decay products $\psi_2$ and $\psi_3$ respectively $ \veq$ are the internal momenta inside decay products, which were assumed to be identical for simplicity.

Moreover $y_{123}$ is the trace of the normalized spin-tensors, corresponding to the spin-angular parts of meson states, and $M(q)$  is a constant string decay amplitude, proportional to the string tension, $M(q)= O(1$ GeV) and for the $L$-wave resonance $\Psi_1$ is proportional to the $p^L$, see appendix . Finally for the width one can write

\be \Gamma(E)= {\rm const}~ p(E)^{2L + 1} |J(p(E))|^2. \label{0.9} \ee
Here $L$ is the angular momentum of the decay products.
So far we are in the realm of the standard hadron decay formalism. Now we take into account that the decay product wave
functions are moving with the velocity $\sqrt {\frac{s - (m_1 + m_2)^2}{s}}$ and hence, their wave functions in momentum space are  Lorentz contracted as shown in (\ref{eq.06}).
To this end we must write $J(p(E))$ in terms of the contracted wave functions, namely, as in (\ref{eq.06}), the wave function, moving with the velocity $v$, can be written as
$\psi_n^{(v)}(\veq) = C_0 \psi_n(\veq_\bot, q_{\|}\sqrt{1-v^2})$. Denoting the total energy $E$, which coincides with the resonance mass at the
resonance center, as $s= E^2$, one can write $C_0 = \sqrt{1-v^2}= \frac{m_2 + m_3}{\sqrt{s}}$.
Therefore the integral in (\ref{0.8}) can be rewritten as
$$J(p) = {\rm const} \int {d^3\veq\Psi_1^{0}(\veq , \vep)\psi_2^{v}(\veq)\psi_3^{v}(\veq)} = $$$$= {\rm const}~ C_0^2 \int {d^2\veq_\bot dq_{\|}\Psi_1^{0}\psi_2(\veq_\bot,q_{\|}\sqrt{1-v^2})\psi_3(\veq_\bot,q_{\|}\sqrt{1-v^2})}=$$\be=
{\rm const}~ C_0 \int d^2\veq_\bot d\kappa \Psi_1^{0} \psi_2(\veq_\bot,\kappa)\psi_3(\veq_\bot,\kappa). \label{0.9} \ee

Here $\kappa = q_{\|}\sqrt{1-v^2}$.
Therefore the decay matrix element is multiplied by $C_0$ and the decay width is multiplied by $C_0^2$.
Summarizing one can write for the two-body decay width of a resonance with account of Lorentz contraction (LC),
firstly in the case of equal masses $m_2 = m_3 = m $, as

\be \Gamma(LC) = C_0^2 \Gamma(0) = \frac{4 m^2}{s} \Gamma(0). \label{10} \ee

Here $\Gamma(0)$ denotes the decay  width without LC dynamics.
In the next sections we shall study the effect of LC in the concrete resonances and compare it with data. We shall also
discuss the case of unequal masses of the decay products and many-body decays.

\section{ Theory of the $\rho$-meson decays with account of Lorentz contraction}

In PDG \cite{25} it is written : ``the determination of the parameters of the $\rho(770)$ is beset with many difficulties because of its large width. In physical region
fits, the line shape does not correspond to a relativistic Breit--Wigner function with a P-wave width, but requires some additional shape parameter."
Indeed in the standard theory with the Lagrangian $L_{eff}= g_{\rho\pi\pi} e_{ijk}\rho^{i\mu}\pi_j \partial_\mu \pi_k$ one obtains the width
\be \Gamma_\rho = \frac{g_{\rho\pi\pi}^2 p^3}{48\pi m_\rho^2}, p= \sqrt{s-4 m_\pi^2},
\label{11}
\ee
the result, which contradicts experimental data. Therefore Gounaris and Sakurai \cite{26} have suggested to modify this
expression, introducing some model dependence of the decay width on the energy,  which is now used in the most data analysis.
The numerous accurate experimental data (see some examples in \cite{27}-\cite{29}) exploit the corrected equation for the
$\Gamma_\rho (s)$, namely, in \cite{27}

\be \Gamma_V (s) = \frac{m_V^2}{s} \frac{p(s)^3}{m_V^3} \Gamma.
\label{12}
\ee

This should be compared with our result in (\ref{10}) for the decay of the $\rho \rightarrow \pi\pi$, where $\Gamma(0)$
refers to the width without the LC, which is proportional to $p(s)^3$ and hence two equations coincide up to the replacement
of $4 m^2$ by $m_V^2$, which is unimportant, since $\Gamma_V$ is a variable numeric parameter. At this point one
meets with an important question: what is the physical wave function of the chiral object-first of all of the pion and how it is connected to the standard quark model solutions. A short discussion in the appendix 2 leads to the answer that the basic chiral wave function is an expansion in the series of the $q-\bar q$ solutions with the leading lowest eigenvalue term. Therefore as explained in the appendix 2 submitting in the form factors or in the decay matrix elements the (lowest) $q- \bar q$ wave function one should consider the corresponding mass $m_{q\bar q}$ in the LC coefficient. Indeed the analysis of the pion form factor in \cite{14} with the help of the LC has shown that the effective pion masses in the $C_0$ should be taken in the nonchiral limit, i.e. around 350 MeV, which leads to the good agreement between the LC pion form factor and lattice and experimental data.
 Turning now to our present problem of the $\rho$ decay width, we take following appendix 2, the pion mass $m= 350$ MeV and obtain numerical agreement of our (\ref{10}) with (\ref{12}). An obvious moral of this action is that
 the masses in the LC coefficient $C_0$ should correspond to the masses of the wave functions used. In the case
 of chiral mesons one meets with the unusual situation when the strong decrease of the chiral object mass coexists with moderate change of the wave function. A simple example- the pion mass of $140$ MeV should correspond to the
 pion radius of around $1.4$ Fm, whereas this radius is $0.6$ Fm, corresponding to the pion mass $350$ MeV.
Note, that as a result the effect of the LC gives the exact reduction coefficient, which can be compared to data and decay theory,
provided such an exact decay theory without arbitrary parameters exists, which is not yet the case for the theory of
strong hadron decays.

We now turn to another example of hadron decay -- the $\pi\pi$ decay of $\rho(1450)$, studied experimentally in \cite{29}, where the authors, following \cite{26},
have used a slightly different from (\ref{12}) parametrization of the width,

\be \Gamma_V (s) = \frac{s}{m^2} \frac{\beta_\pi(s)^3}{\beta_\pi(m)^3} \Gamma = \frac{m p^3(s)}{\sqrt{s p^3(m)}}\Gamma,
\label{13}
\ee
where $\beta_\pi(s)= \sqrt{1-4m_\pi^2/s}$.
One can see the factor $\sqrt{s/m^2}$ difference between (\ref{12}) and (\ref{13}), but from the point of view of our theory the result of (\ref{12}) is preferable.
It is clear that numerous experimental results on both $\rho(770)$ and $\rho(1450)$ require a more detailed quantitative analysis using the present approach with modified width equations, nevertheless the first comparison discussed above can be taken as a support of our LC formalism.

\section{ Extensions and discussion }

We shall discuss below third possible extensions of the above formalism: (A) decays to two unequal mass hadrons; (B) decays to one hadron and an elementary object
(e.g. $\gamma$); (C) decays to tree or more hadrons.

  (A) Till now we discussed strong hadron decay to two equal mass mesons. This definition presupposes the strong interaction
decay matrix element, containing an integral of decay product wave functions as in (\ref{0.9}). We now turn to the case
of unequal masses $m_2, m_3$, where the particles $2,3$ move with velocities $v_2,v_3$, where $v_i= \frac{p}{\sqrt{m_i^2 + p^2}}$. Here $p$ is the momentum of particle in the decay process. As it is easy to see
in (\ref{0.9}) one obtains in the decay matrix element the factor $K(LC)= C_0(v_1) C_0(v_2) I_{23}$, where $I_{23}= \int dq_{\|}\psi_2(q_{\|})\psi_3(q_{\|})$.
To proceed we assume for the decay product wave functions the Gaussian form, so that for the longitudinal part of the wave functions one has
$\psi_i(q_{\|})= N_i \exp\left(-\frac{q_{\|}^2}{\chi_i^2}\right)$, which yields for the integral

\be
I_{23}= \sqrt{\frac{2}{\frac{m_2^2 \chi_3}{(p^2 +m_2^2)\chi_2} + \frac{m_3^2 \chi_3}{(p^2 + m_3^2)\chi_2}}}.
\label{14}
\ee
It is clear that $\chi_i^{-1}$ plays the role of the effective radius of the state $r_i$, which we shall use in what follows.
As a result one can define the LC coefficient $K(LC)$, which is equal to
\be
K(LC)= \sqrt2 \frac {C_0(v_2)C_0(v_3)}{\sqrt{C_0(v_2)^2 r_2/r_3 + C_0(v_3)^2 r_3/r_2}}.
\label{15}
\ee

From (\ref{15}) one can see that in the case when the meson 2 is more light ($m_2 << m_3 $) and/or more narrow $r_2 <<r_3$,  its contraction coefficient enters in the final
answer $K(LC)= C_0(v_2)\sqrt{r_2/r_3}$.
To understand better the situation with unequal masses we can compare two decays $\rho(1450)\rightarrow \omega\pi$,
which we define as the decay $(1)$, and $\rho(1450)\rightarrow \pi\pi$ as the decay $(2)$.
Assuming decay constants as equal and all difference only due $P$-wave momenta and the LC coefficients, one obtains

\be \frac{\Gamma_1}{\Gamma_2} = \frac{p(\omega\pi)^3 K_1(LC)^2}{p(\pi\pi)^3 K_2(LC)^2} = 2\frac{p_1 r_2}{p_2 r_1}= 1.44 \frac{r_{\pi}}{r_{\omega}} \label{16} \ee
which is O(1) and agrees with almost equal widths of decays $(1)$ and $(2)$.

  (B) Consider e.g. the process $h(1) \rightarrow h(2) + \gamma$ where $\gamma$ has no internal structure and the corresponding wave function. It is clear that
the decay matrix element has the structure in the c.m. of $h(1)$
can be written in full analogy with the form factor, written in (\ref{18}) of \cite{14}

\be J(123) = {\rm const} \int d^3 q \psi_1^{ 0 }(\veq) \psi_2^{Q}\left(\veq +\veQ \frac{\omega^{''}}{\omega' + \omega^{''}}\right),
\label{17}
\ee
where $\omega',\omega^{''}$ are the stationary values of relativistic energies of two particles  in the path integral (see \cite{9,10,11}) and in this matrix element
$\gamma$ is emitted by the particle'. In (\ref{17}) the upper index $(Q)$ in
$\psi$ denotes the momentum $Q$ of the hadron and, as in (\ref{0.9}), one obtains an extra factor $C_0(Q)$. Then introducing $\kappa$, as in Eq.~(\ref{0.9}), this factor is cancelled.
As a result one is left with the standard expression with exception that $\veQ$ in the argument of $\psi_2$ is multiplied by the factor $\sqrt{1- v_Q^2}$, which strongly
weakens the $Q$ dependence at large $Q$.
These properties of the $\gamma$ transitions are applicable  to all decays, including elementary objects without internal structure.

\section{Conclusions and an outlook}

   Till now we discussed the two-body decays of different types. It is clear that the same approach can be applied to more general cases. As an example we now consider strong decays to three and more hadrons, e.g. $(1)\rightarrow (2,3,4)$, and assuming again the string-breaking mechanism, the decay matrix element,
as in \cite{17}, will be proportional to the factor,

\be J(1234) = \int d^3 q \Psi_1(q,p_2,p_3,p_4)\psi_2^{p_2}(q)\psi_3^{p_3}(q)\psi_4^{p_4}(q),
\label{18}
\ee
where $p_i$ in  $\psi_i^{p_i}$ is the momentum of the hadron $i$ , $\vep_1 + \vep_2 + \vep_3 = 0$, so that the hadron
wave functions acquire the factors $C_0(v_i)$ and the longitudinal momenta of $\veq$ in $\psi_i(\veq)$ are multiplied
by the same factors $C_0(v_i)$, different in general for all $i$. This creates a rather unusual distribution in the
Dalitz plane which will be studied elsewhere.

So far we have investigated only the most general and simple consequences of the LC for the wave functions in the case
of hadron decays. We should stress that our analysis refers to the strong decays of hadrons, when the decay is assumed
to proceed in the nonperturbative  way as in the string decay mechanism, so that in the decay matrix element participate the hadron wave functions as the whole objects,
consisting of the quarks and gluons, connected by instantaneous strong interaction producing a string. Therefore in the decay of the string one obtains immediately again
two string objects with their full wave functions and not separate quark and gluons, as it would be in the perturbative approach.
From this point of view the use of the LC mechanism for the strong hadron decays seems to be well founded and the good  agreement of the reduction coefficient $C_0^2$ with the well and long proved experimentally in $\rho(770)$ and $\rho(1450)$ decays gives additional support for it.

We have not discussed all consequences of this LC formalism,
which works also reasonably well in the form factors of mesons \cite{15} and baryons \cite{16}, and can in principle be
used in all reactions, where hadron wave functions enter explicitly. To proceed further one needs to develop an ``anatomy" of the decay and in general
of the hadron exchange and creation processes, which is in progress.

The author is grateful to A.M.Badalian for useful discussions and advices.
This work is supported by the Russian Science Foundation (RSF) in the framework of the scientific project,Grant 16-12-10414.

  \vspace{2cm}

{\bf Appendix A1.} {\bf The $L$ dependence of the decay matrix element}  \\

 \setcounter{equation}{0} \def\theequation{A1.\arabic{equation}}

Starting from the (\ref{0.8}) one can write the decay matrix element $J_L(p)$, when the string extending from $x_Q$ to $x_{\bar Q}$ decays at point $x$ into two strings with the appropriate wave functions $\psi_2,\psi_3$
\be
J_L(p)= \int d^3x d^3(x_Q-x_{\bar Q}) \exp{ip(x_Q- x_{\bar Q})} \Psi_1^{(L)}(x_Q- x_{\bar Q}) \psi_2(x_Q -x) \psi_3(x-x_{\bar Q}).
\label{eq.A1.1} \ee
Going over into the momentum space one obtains
\be
J_L(p)= \int d^3q \Psi_1^{(L)}(p+q) \psi_2(q) \psi_3(q)
\label{eq.A1.2} \ee
Here $\Psi_1^L(p) ~ p^L$ and for $L=1$ one obtains the linear dependence on $p$.
One should stress the important property of (\ref{0.8a}) and (\ref{eq.A1.1}) - both hadrons 2 and 3 have the same internal momentum $q$ which leads finally to the effects of Lorentz contraction in the decay width, while in absence of string breaking at the point $x$ both momenta $q$ and $q'$ in $\psi_2(q)$ and $\psi_3(q')$ are independent, they are integrated separately and no contraction effects are visible.

\vspace{1cm}

{\bf Appendix A2.} {\bf The pion wave function}\\

 \setcounter{equation}{0} \def\theequation{A2.\arabic{equation}}

The problem of chiral wave functions was studied in a series of papers during last 20 years (see \cite{30} and refs therein) using the Chiral Confining Lagrangian (CCL) where CSB is directly connected with confinement and all known chiral relations like GMOR \cite{31} are directly deduced from CCL.
The main outcome of the CCL for the pion Green's function in \cite{30} is that it can be expanded as a series of $q\bar q$ wave functions $(\phi_n(x)$ while the pion mass is strongly shifted from the set $m_n$ and is defined by the vacuum condensate $\lan\bar q q\ran$ as it is seen in GMOR relations.
As a result in \cite{30} the pion Green's function was obtained in the form
$ G(k)= \frac{\Psi(k)}{(k^2 + m_{\pi}^2) \Phi(k)}$, where both $\Psi(k),\Phi(k)$ are expressed via $\phi_n(x)$
only and the lowest eigenfunction $\phi_0(x)$ gives the dominant contribution. In a similar way the basic chiral
parameters $\lan \bar q q \ran$ and $f_\pi$ are expressed in \cite{30} as a sum of $\phi_n$ contributions with the dominant role of $\phi_0$. Summarizing one can conclude that the pion is well described by the standard $q \bar q$ nonchiral wave functions except for its mass which is strongly decreased by the vacuum effects.

\end{document}